\documentclass[a4paper]{aa}

\usepackage{graphicx}
\usepackage{txfonts}
\usepackage{natbib}
\usepackage{aalongtable}
\bibpunct{(}{)}{;}{a}{}{,}

\begin{document}

\title{GaBoDS: The Garching-Bonn Deep Survey }
\subtitle{III. Lyman-Break Galaxies in the Chandra Deep Field South\thanks{Based on
    observations made with ESO Telescopes at the La Silla Observatory.}}
\titlerunning{Lyman-break galaxies in the CDFS}

   \author{H. Hildebrandt
          \inst{1}
          \and
          D.~J. Bomans\inst{2}
          \and
          T. Erben\inst{1}
          \and
          P. Schneider\inst{1}
          \and
          M. Schirmer\inst{3}
          \and
          O. Czoske\inst{1}
          \and
          J.~P. Dietrich\inst{1}
          \and
          T. Schrabback\inst{1}
          \and
          P. Simon\inst{1}
          \and
          R.~J. Dettmar\inst{2}
          \and
          L. Haberzettl\inst{2}
          \and
          M. Hetterscheidt\inst{1}
          \and
          O. Cordes\inst{1}
          }

   \offprints{H. Hildebrandt \email{hendrik@astro.uni-bonn.de}}

   \institute{Institut f\"ur Astrophysik und extraterrestrische
     Forschung, Universit\"at Bonn, Auf dem H\"ugel 71, D-53121 Bonn, Germany\\
         \and
             Astronomisches Institut der Ruhr-Universit\"at-Bochum,
             Universit\"atsstr. 150, D-44780 Bochum, Germany\\
         \and
             Isaac Newton Group of Telescopes, Apartado de correos 321, 38700 Santa Cruz de La Palma, Tenerife, Spain
             }
   \date{Received ; accepted }
   
   \abstract{We present first results of our search for high-redshift galaxies
     in deep CCD mosaic images. As a pilot study for a larger survey, very
     deep images of the Chandra Deep Field South (CDFS), taken with
     WFI@MPG/ESO2.2m, are used to select large samples of 1070 $U$-band and
     565 $B$-band dropouts with the Lyman-break method. The data of these
     Lyman-break galaxies are made public as an electronic table. These
     objects are good candidates for galaxies at $z\sim3$ and $z\sim4$ which
     is supported by their photometric redshifts. The distributions of
     apparent magnitudes and the clustering properties of the two populations
     are analysed, and they show good agreement to earlier studies. We see no
     evolution in the comoving clustering scale length from $z\sim3$ to
     $z\sim4$. The techniques presented here will be applied to a much larger
     sample of $U$-dropouts from the whole survey in near future.
   
   \keywords{Galaxies: photometry -- Galaxies: high-redshift -- Surveys
            }
   }

   \maketitle

\section{Introduction}
\label{sec:introduction}
In order to constrain models of structure formation and to investigate the
star formation history of the universe, large samples of galaxies at
high redshift are needed. The clustering properties on large scales at these
high redshifts can only be studied with contiguous fields of considerable
size. Furthermore, to overcome cosmic variance, different lines of sight
should be probed. The Lyman-break technique is an efficient method to select
galaxies at high redshift from multi-colour optical data.

The largest survey of Lyman-break galaxies (LBGs) at $z\sim3$ to date
\citep{2003ApJ...592..728S} covers 0.38 square degrees in 17 widely separated
fields yielding more than 2000 LBG candidates of which 940 have been confirmed
spectroscopically.  On a sub-area, this group has published 244 $G$-dropouts
(48 spectroscopically confirmed), candidates for $z\sim4$ galaxies
\citep{1999ApJ...519....1S}.  \citet{2003A&A...409..835F} published results
from the Canada-France deep fields identifying $\sim 1300$ $U$-dropouts on a
shallower but larger field.  Very recently, \citet{2004ApJ...611..660O}
obtained a sample of $\sim2000$ $B$-dropouts in deep Suprime-Cam imaging and
observed 85 of them spectroscopically.

In this paper we investigate large samples of $U$- and $B$-dropouts in very
deep wide-field images of the CDFS. This investigation on one field is a pilot
study for a much larger survey, the ESO Deep-Public-Survey (DPS), of LBGs on
mosaic CCD data. Special attention is paid to the careful selection of
candidates and the comparison with other successful studies of LBGs. While
still smaller than some other samples because of the limited area, our LBG
population will grow significantly in the near future with the analysis of the
whole survey. Here the methods that will be applied to the complete dataset
are presented and evaluated.

In Sect.~\ref{sec:Observations_and_data_reduction} the observations, the data
reduction, and the catalogue extraction are described.  Sect.
\ref{sec:sample_selection} deals with the photometric selection of dropouts in
our data. After that the properties of the two dropout samples are presented
in Sect.~\ref{sec:properties-samples}. Photometric redshift estimates, the
distributions of apparent magnitudes, and the clustering properties are shown
there. Concluding remarks and an outlook are given in Sect.
\ref{sec:conclusions}.

\section{Observations and data reduction}
\label{sec:Observations_and_data_reduction}

\subsection{Observations}
The Chandra Deep Field South ($\alpha=03^\mathrm{h}\,32^\mathrm{m}\,29^\mathrm{s}$,
$\delta=-27\degr\,48\arcmin\,47\arcsec$) was observed with the WFI@MPG/ESO2.2m
for several programmes. Data were taken for the GOODS project
\citep{2004ApJ...600L..93G}, the COMBO-17 survey \citep{2004A&A...421..913W},
and the ESO-Imaging-Survey (EIS) \citep{2001A&A...379..740A}. All these data
are available from the ESO archive. Erben et al. (ESO Press Photos 02a-d/03)
have produced very deep images in $BVR$ with a field-of-view of $34'\times33'$
using the Bonn WFI reduction pipeline (\citeauthor{2003A&A...407..869S}
\citeyear{2003A&A...407..869S}; \citeauthor{2005astro.ph..1144E}
\citeyear{2005astro.ph..1144E}).  Additionally, $U$- and $I$-band images were
published by the EIS-team \citep{2001A&A...379..740A}.  Their properties are
summarised in Table~\ref{tab:CDFS}.\footnote{If not otherwise specified we use
  Vega magnitudes in this paper.}

\begin{table*}
\caption{Properties of the CDFS WFI-data. The limiting magnitudes in
  columns 3 and 4 are calculated with equation \ref{equ:mag_lim}.}
\label{tab:CDFS}
\centering
\begin{tabular}{c c c c c c c c}
\hline\hline
Band & ESO-Id & exposure & $3\sigma$ limits in a $2''$ diam. aperture & $1\sigma$
limits in $2\times$ FWHM diam.& AB         &
FWHM & source      \\
     & & time [s] & (Vega mags) & (Vega mags)      & correction & [$\arcsec$] & \\
\hline
$U$ & U/50   & 43\,600 & 25.6 & 26.8 & 0.9    & 1.07 & EIS\\
$B$ & B/99   & 57\,000 & 28.0 & 29.2 & $-$0.1 & 0.99 & Bonn/GaBoDS\\
$V$ & V/89   & 56\,000 & 27.5 & 28.7 & 0.0    & 0.93 & Bonn/GaBoDS\\
$R$ & Rc/162 & 57\,100 & 27.6 & 28.7 & 0.2    & 0.81 & Bonn/GaBoDS\\
$I$ & Ic/lwp & 26\,900 & 25.1 & 26.3 & 0.5    & 0.95 & EIS\\
\hline
\end{tabular}
\end{table*}

The $BVR$ images were coadded with \emph{drizzle} \citep{2002PASP..114..144F}.
The astrometric calibration was done with respect to the USNO-A2.0
\citep{1998yCat.1252....0M} and the photometric calibration is based on the
COMBO-17 CDFS data \citep{2004A&A...421..913W}.

The properties of the $U$- and $I$-band images from EIS are described in
detail in \citet{2001A&A...379..740A}. The astrometric solution for these
images is recalculated on the basis of our $R$-band catalogue.

\subsection{Image preparation and catalogue extraction}
Since the EIS images come from a different reduction pipeline it is necessary
to resample them again to exactly the same output grid with the same centre
coordinates in order to use the dual image mode of \emph{SExtractor}
\citep{1996A&AS..117..393B} described below. This is done by \emph{SWarp}
\citep{2003SWarp} which minimises the introduction of additional noise by
applying a reverse mapping technique combined with an advanced kernel
function (Lanczos-3).

The catalogues are created using \emph{SExtractor} in dual-image mode. In this
mode objects are detected and their shapes are measured on the $R$-band image.
The flux in the other bands is measured on the corresponding images at the
positions derived from the $R$-band. The $R$-band is chosen as the detection
image since it is very deep, has very good seeing, and the targeted LBGs are
comparatively bright in this band. An object is detected in the $R$-band if
the flux in five adjacent pixels exceeds the standard deviation of the local
sky background fluctuations by a factor of three. This conservative criterion
is chosen because the handling of dropouts in blue bands requires clear
detections in redder bands.

To account for the different seeing properties of the images, aperture
magnitudes are used with the size of the aperture in one band scaled to the
seeing of that image (diameter of the aperture $=2\times$FWHM) when colours
are measured.  This approach is justified for the investigation of LBGs since
these objects are usually not resolved in ground-based images. Thus, our
approach delivers correct colours as long as the seeing is not too different
in the images used (see Table~\ref{tab:CDFS}).
When magnitudes are cited in the following, the \emph{SExtractor} parameter
MAG\_AUTO is used which corresponds to flexible elliptical apertures described
in \citet{1980ApJS...43..305K}. The aperture magnitudes are used only for
colour estimation.

When objects are detected in the $R$-band image and the flux is measured in
the other bands, it is necessary to separate detected from non-detected objects
in the bands different from the $R$-band. For that purpose limiting magnitudes
for the apertures defined above are calculated:

\begin{equation}
  \label{equ:mag_lim}
  mag_{\mathrm{lim}}=ZP-2.5 \log \left(\sqrt{N_{\mathrm{pix}}}\cdot\sigma\right) \; .
\end{equation}
$ZP$ is the photometric zeropoint of the image, $N_{\mathrm{pix}}$ is the
number of pixels in the aperture, and $\sigma$ gives the global RMS
pixel-to-pixel fluctuations of the sky background in the image considered. In
Table~\ref{tab:CDFS} two different limiting magnitudes are given for every
image, $3\sigma$ limits in a $2''$ diameter aperture and the $1\sigma$ limits
in an aperture with $2\times$FWHM diameter. The latter are used to set a
lower/upper limit to the colour index of objects that are not detected in one
band.

Our final catalogue contains $\sim\!57\,000$ $R$-band detected objects of which
$\sim\!10\,000$ have no significant flux in $U$, $\sim\!300$ are not detected in
$B$, $<\!100$ are not detected in $V$ (mostly $R$-band image defects), and
$\sim\!4\,500$ are not detected in $I$ (due to the shallower depth of this
image). No star-galaxy separation is performed.  LBGs are of
such small apparent size that a considerable fraction of them would possibly
be misclassified as stars and rejected if this was done.

\section{Sample selection}
\label{sec:sample_selection}
Whenever selecting a sub-population from a large catalogue, attention must be
paid to maximise completeness and efficiency. Often these goals are working in
opposite directions and a good compromise must be chosen. While the real
efficiency can usually be quantified with spectroscopic data at hand, the
completeness is hard to quantify.  Even defining completeness and efficiency
can be somewhat ambiguous in this context as we show in the following. Here we
are searching for high-redshift galaxies, which means that stars or
low-redshift interlopers can contaminate the catalogues, thus reducing the
efficiency.  The case for intermediate redshift ($1\!<\!z\!<\!2$) galaxies is
more difficult.  In principle, we are highly interested in these objects since
few are known up to now \citep[see][]{2004ApJ...604..534S}. But for the
clustering analysis and obtaining luminosity functions in redshift slices,
these objects are also contaminants and should be separated. To guarantee a
reasonable efficiency, model galaxies' colours are investigated for
high-redshift galaxies with a pronounced Lyman-break as well as for
low-redshift ellipticals that are nearby in colour space. Furthermore, our
selection follows other successful studies of LBGs cited above.

Completeness, however, is a different issue. If our goal was to select every
galaxy at e.g. $z\sim3$ in our data we would not be very complete with the
method described below. In fact we are searching for LBGs which are easy to
detect because of their pronounced Lyman-break. More dusty galaxies are much
harder to separate from low-redshift objects, and if they are common at high
redshift we will miss a lot of them with our selection criteria. Today it is
known that LBGs are common at high redshift and not rare objects, representing
a considerable fraction of the total galaxy population at these epochs
\citep{2002ARA&A..40..579G}.

\subsection{Colours of high-redshift galaxies}
\begin{figure}
\resizebox{\hsize}{!}{\includegraphics{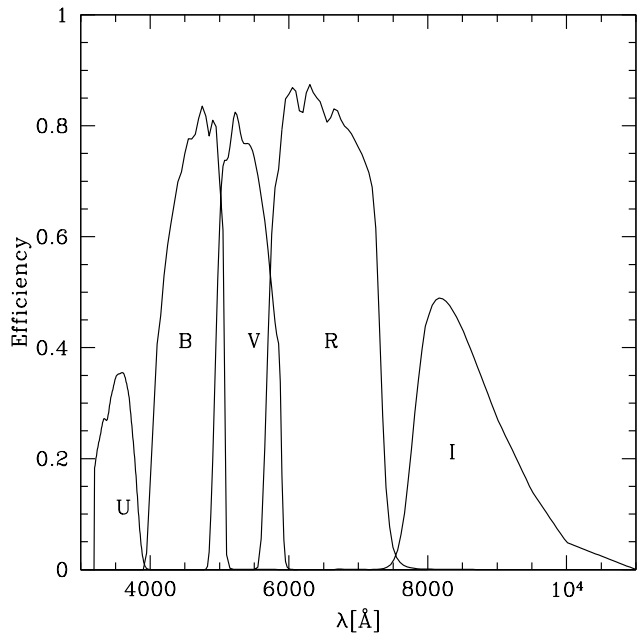}}
\caption{\label{fig:WFI_filter_curves_plus_CCD}Instrumental response
  of WFI in the different filters.}
\end{figure}

The publicly available photometric redshift code \emph{Hyperz}
\citep{2000A&A...363..476B} is used to estimate the colours of high-redshift
galaxies. Template spectra from the library of \citet{1993ApJ...405..538B} are
taken and convolved with the instrumental response of the WFI (see Fig.~\ref{fig:WFI_filter_curves_plus_CCD}).  The spectral energy distribution (SED)
of a galaxy with constant star formation rate (spectral type: Im) is chosen
which has a pronounced Lyman-break. Different amounts of reddening are taken
into account by applying the dust extinction law of
\citet{2000ApJ...533..682C}. The opacity of the intergalactic medium is
included by applying the estimates from \citet{1995ApJ...441...18M}.

\begin{figure*}
\centering
\includegraphics[width=17cm]{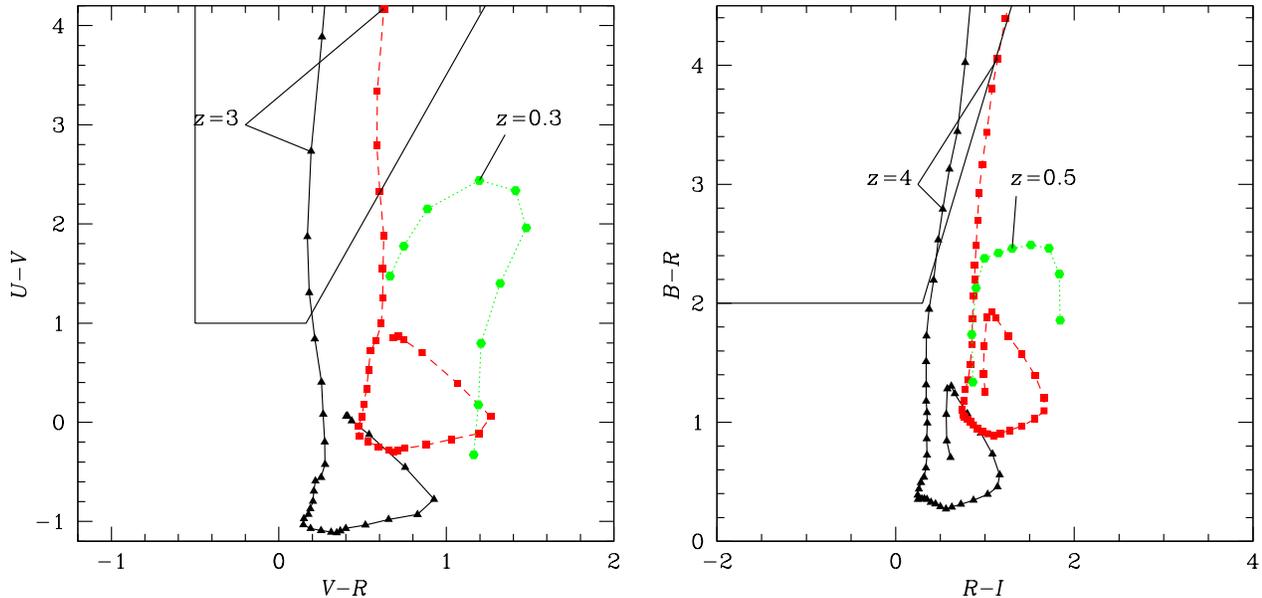}
\caption{\label{fig:models}Colours of model galaxies in the
  $U-V$ vs. $V-R$ two-colour diagramme used for $U$-dropout selection
  (\emph{left}) and in the $B-R$ vs. $R-I$ two-colour diagramme for $B$ dropout
  selection (\emph{right}).  The solid lines represent galaxies (spectral type
  Im) with no dust reddening, the dashed lines represent galaxies (spectral
  type Im) with an extinction in the visual of $A_V=1.5$~mag, and the dotted
  lines represent elliptical galaxies (spectral type E) at low redshift. The
  points correspond to intervals of $\Delta z=0.1$.  The boxes define our
  selection boundaries for high-$z$ galaxy candidates.}
\end{figure*}

Furthermore, the colours of elliptical galaxies at low redshift are calculated
in an identical way in order to estimate their contamination of our samples of
LBGs. In Fig.~\ref{fig:models} the colour distributions of the
model galaxies are shown.

\subsection{Selection of Candidates}
\label{sec:selection-candidates}
\begin{figure}
\resizebox{\hsize}{!}{\includegraphics{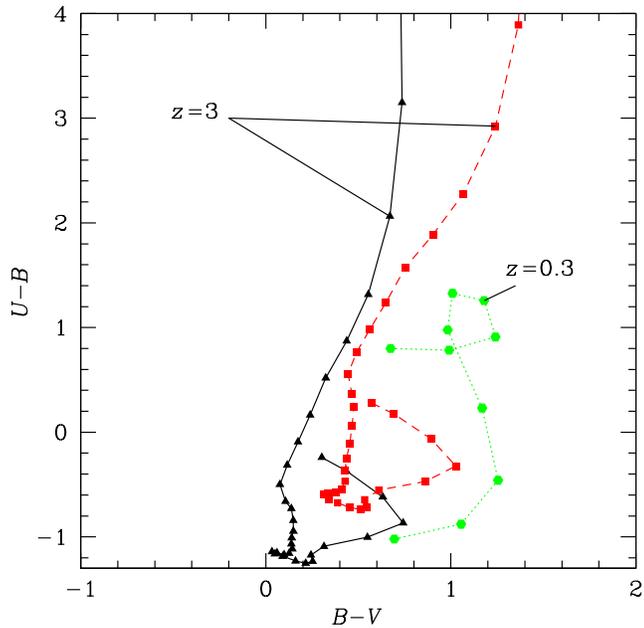}}
\caption{\label{fig:Up_B_V_models}Colours of model galaxies. The
  solid line represents galaxies (spectral type Im) with no dust reddening,
  the dashed line represents galaxies (spectral type Im) with an extinction in
  the visual of $A_V=1.5$~mag, and the dotted lines represent elliptical
  galaxies (spectral type E) at low redshift. The points correspond to
  intervals of $\Delta z=0.1$.  Here the effect of overlapping filters can be
  seen, resulting in slightly diagonal tracks which make it difficult to
  choose a selection box.}
\end{figure}

We based our selection criteria for high-redshift objects on the predicted
colours of model galaxies as outlined above. Given our filter set (see Fig.~\ref{fig:WFI_filter_curves_plus_CCD}) and the data quality in the different
bands, objects with $z\sim3$ are selected most efficiently in a $U-V$, $V-R$
colour-colour diagramme. More distant galaxies at $z\sim4$ are preferentially
picked up in the $B-R$, $R-I$ space. In principle those populations can also
be selected in $U-B$, $B-V$ and $B-V$, $V-R$ space respectively.  However, due
to the significant wavelength overlaps between the $B$, $V$ and $R$ filters,
and due to the small gap between $U$ and $B$, an efficient discrimination
between galaxies with and without a pronounced Lyman-break is not possible in
such diagrammes. This can be seen in Fig.~\ref{fig:Up_B_V_models} where the
redshift tracks run more diagonal than in Fig.~\ref{fig:models} due to the
fact that an object that `drops out' from the $U$-band completely becomes
already significantly fainter in the $B$-band. For the same reason a search
for $V$-dropouts in our data is difficult, although our very deep wide-field
$V$-band is predestined for such a project. Deep infrared data from the GOODS
project with ISAAC@VLT \citep{2004ApJ...600L..93G} are available for the
innermost part of our field and will help in searching for $V$-dropouts (see
Sect.~\ref{sec:conclusions}).

The selection must always be a compromise between completeness and efficiency.
Galaxies that are too red in $V-R$ cannot be included in the $z\sim3$
selection box, for example, if one wants to avoid contamination by
low-redshift elliptical galaxies. The same is true for the $z\sim4$ sample.
Photometric errors will scatter the data-points of faint galaxies around in the
two-colour-diagrammes so that it is not possible to predict a precise redshift
distribution of the samples. Furthermore, the redshift distribution will
change with intrinsic spectral shape because of complex selection effects.

\begin{figure*}
\includegraphics[width=17cm]{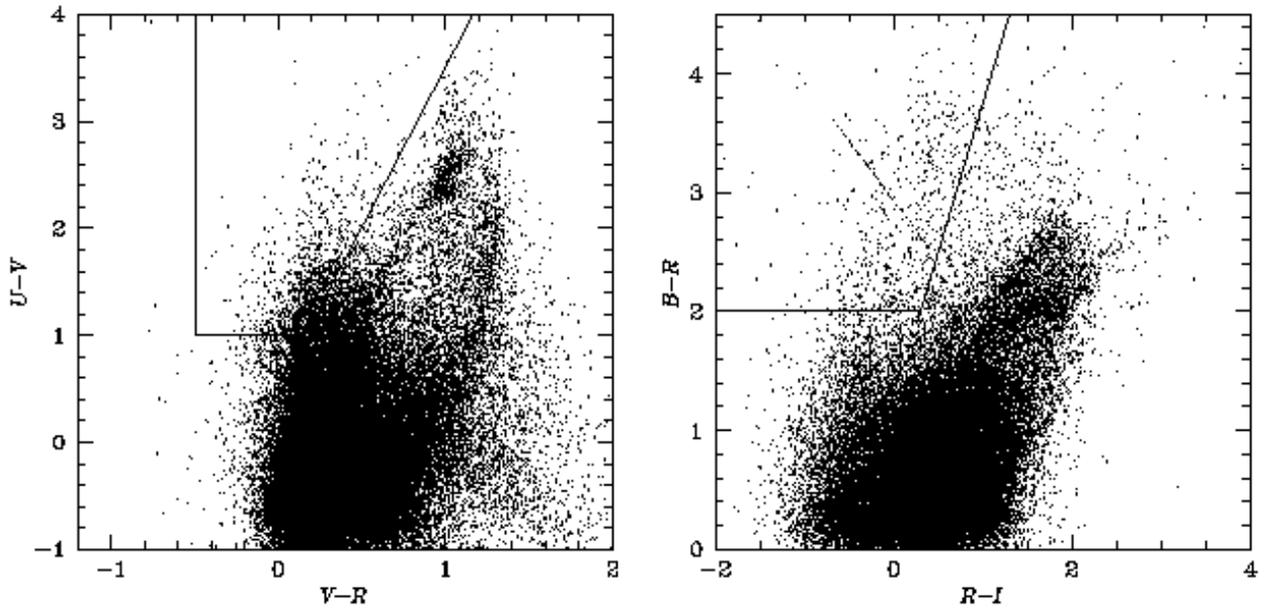}
\caption{\label{fig:objects}$(U-V)$ vs. $(V-R)$ (\emph{left}) and $B-R$
  vs. $R-I$ (\emph{right}) colours of galaxies in the CDFS WFI catalogue. The
  boxes represent the selection criteria given in equation
  \ref{equ:U_dropout_selection} (\emph{left}) and
  \ref{equ:B_dropout_selection} (\emph{right}). The spike running from the
  lower right to the upper left inside the $B$-dropout selection box
  (\emph{right}) is an artifact due to the inferior depth of the $I$-band image
  (see text). It does not affect the efficiency of the dropout selection.}
\end{figure*}

Based on these considerations the following selection criteria are
chosen (see Fig.~\ref{fig:models}). For the $U$-dropout selection,
\begin{eqnarray}
  \label{equ:U_dropout_selection}
  1 &\le&(U-V) \; , \nonumber \\
  -0.5 &\le& (V-R)\le1.5 \; ,\\
  3\cdot(V-R) &\le& (U-V)-0.5 \; , \nonumber
\end{eqnarray}
and for the $B$-dropout selection,
\begin{eqnarray}
  \label{equ:B_dropout_selection}
  2 &\le&(B-R) \; , \nonumber \\
  && (R-I)\le1.5 \; ,\\
  2.5\cdot(R-I) &\le& (B-R)-1.25 \; . \nonumber
\end{eqnarray}
Applied to our catalogues, we get 1167 $z\sim3$ $U$-dropout candidates and 613
$z\sim4$ $B$-dropout candidates (see Fig.~\ref{fig:objects}). All $U$-dropout
candidates are detected in the $B$-, $V$-, and $R$-band images, so that their
colour selection is not influenced by the depth of these images. 101 of them
are not detected in $I$ being fainter than $I=26.3$, the detection limit in
that band. The colour selection of the intrinsically fainter $B$-dropouts is
also not seriously influenced by that effect, although 172 of them are not
detected in $I$ (this is the reason for the spike running from the lower right
to the upper left in the selection box of Fig.~\ref{fig:objects}).  Their
$(R-I)$ colour is an upper limit. There are, however, some objects that lie to
the right of our selection box, which could have bluer $(R-I)$ colours. So,
efficiency is not affected while completeness suffers from the lower depth of
the $I$-band image.

\begin{figure*}
\centering
\includegraphics[width=17cm]{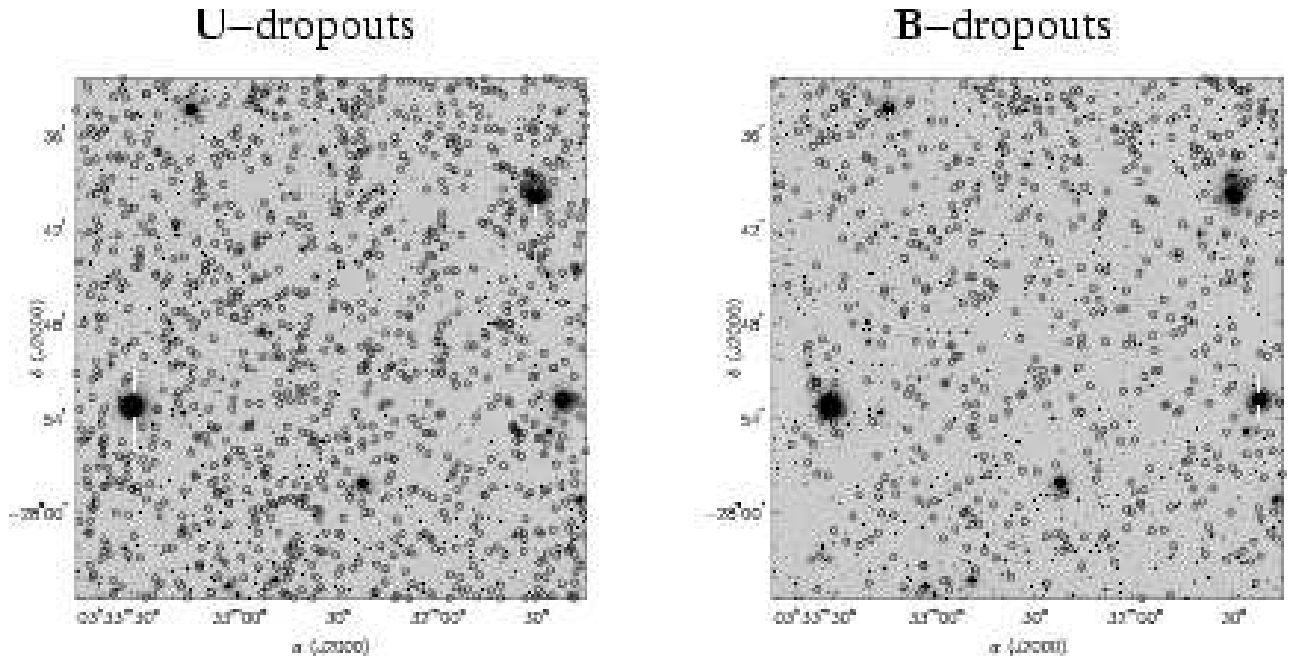}
\caption{\label{fig:spatial_dist}Spatial distribution of our $U$-dropouts
  (\emph{left}) and our $B$-dropouts (\emph{right}).}
\end{figure*}

Thumbnail pictures in the five WFI bands are created for all selected objects.
Some examples are shown in Fig.~\ref{fig:GOODS_thumbnails} and
\ref{fig:GEMS_thumbnails}. Every candidate is checked by eye and some spurious
detections like bad pixels, cosmic rays, reflections, or other image defects
are rejected. After that our catalogues still contain 1070 $U$-dropouts and
565 $B$-dropouts. Our dropout catalogues are freely available to the
scientific community in the electronic version of A\&A (Tables~\ref{tab:cat_U}
and \ref{tab:cat_B})\footnote{Tables~\ref{tab:cat_U} and \ref{tab:cat_B} are
  also available in electronic form at the CDS via anonymous ftp to
  cdsarc.u-strasbg.fr (130.79.128.5) or via
  http://cdsweb.u-strasbg.fr/cgi-bin/qcat?J/A+A/}. The spatial distribution of
the two samples is shown in Fig.~\ref{fig:spatial_dist}.

\subsection{Observations with other telescopes}
\begin{figure*}
\includegraphics[width=17cm]{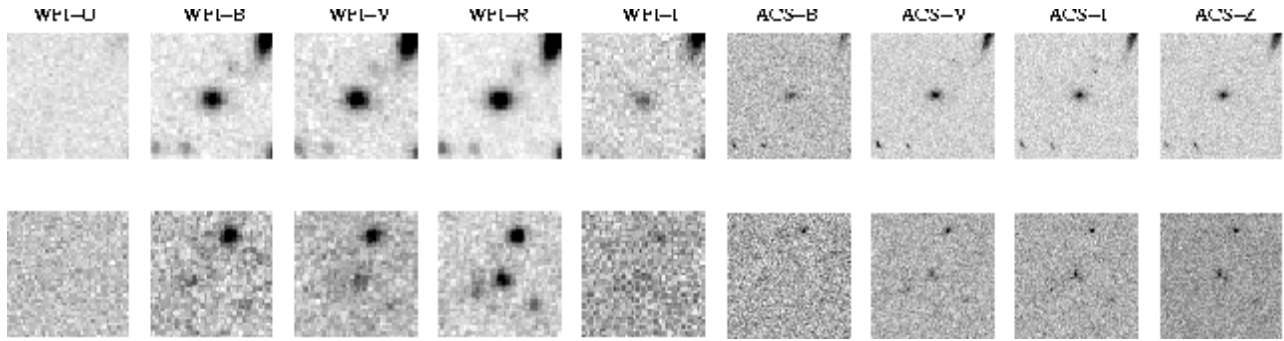}
\caption{\label{fig:GOODS_thumbnails}Examples of a $U$-dropout (upper
row) and a $B$-dropout (lower row) in the GOODS area. Thumbnail
pictures in the WFI-$UBVRI$ and ACS-$BVIZ$ filters (from left to right) of
size $10''\times 10'' $.}
\end{figure*}

\begin{figure*}
\includegraphics[width=17cm]{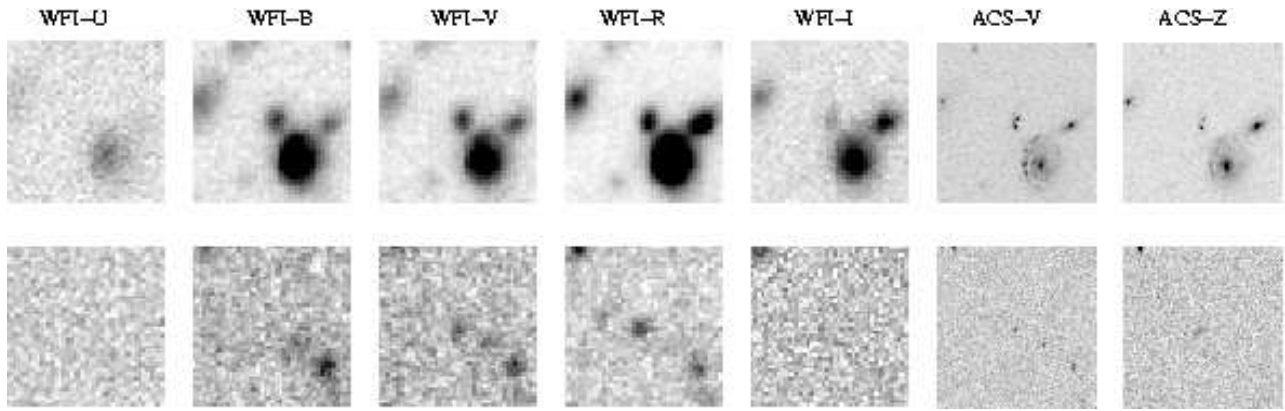}
\caption{\label{fig:GEMS_thumbnails}Examples of a $U$-dropout (upper
row) and a $B$-dropout (lower row) in the GEMS area. Thumbnail
pictures in the WFI-$UBVRI$ and ACS-$VZ$ filters (from left to right) of size
$10'' \times 10'' $. In the ACS images the irregular nature of this
$U$-dropout is clearly revealed, while the $B$-dropout is barely visible in
the WFI-$I$-band because of the limited depth of this image.}
\end{figure*}

The whole WFI field is covered with data from the Advanced Camera for Surveys
(ACS) on board Hubble Space Telescope (HST). The inner part of our
$34'\times33'$ field was observed for the GOODS programme
\citep{2004ApJ...600L..93G} in the four bands $BVIZ$ (F435W, F606W, F775W and
F850LP) and the outer regions were observed for the GEMS project
\citep{2004ApJS..152..163R} in $V$ and $Z$ (F606W and F850LP).  From these
data, thumbnail pictures for nearly every dropout candidate are created and
examples are also shown in Fig.~\ref{fig:GOODS_thumbnails} and
\ref{fig:GEMS_thumbnails}.

Since the GOODS data cover the $BVIZ$ filters, they are suited to check our
$B$-dropout selection criteria. The ACS thumbnails of all the $B$-dropouts
inside the GOODS area are checked by eye for contamination by stars (point-like
objects) or objects that are clearly visible in the ACS $B$-band image. Seven
out of 66 objects are classified as possible contaminants. Thus, we roughly
estimate the efficiency of our $B$-dropout selection to $\sim 90\%$.

Six of our $U$-dropouts have been observed spectroscopically in the VVDS
\citep{2004astro-ph...0403628} and one of them also for the GOODS programme
\citep{2004astro-ph...0406591}. Three of them are at $z>3$, while the other
three are low-redshift interlopers.  This is not surprising since all of these
objects are quite bright ($R\sim23$) and contamination plays an important role
in these magnitude ranges \citep[see][]{2003ApJ...592..728S}. The redshift of
one of those interlopers is not yet determined unambiguously, estimates range
from $z=0.22$ (GOODS) to $z=0.64$ (VVDS). None of our $B$-dropouts have been
observed spectroscopically so far.

\section{Properties of the samples}
\label{sec:properties-samples}
In this section we test our selection criteria in detail by analysing and
comparing our LBG samples against those of other studies.

\subsection{Photometric redshift distributions}
\begin{figure*}
%\sidecaption
\includegraphics[width=12cm]{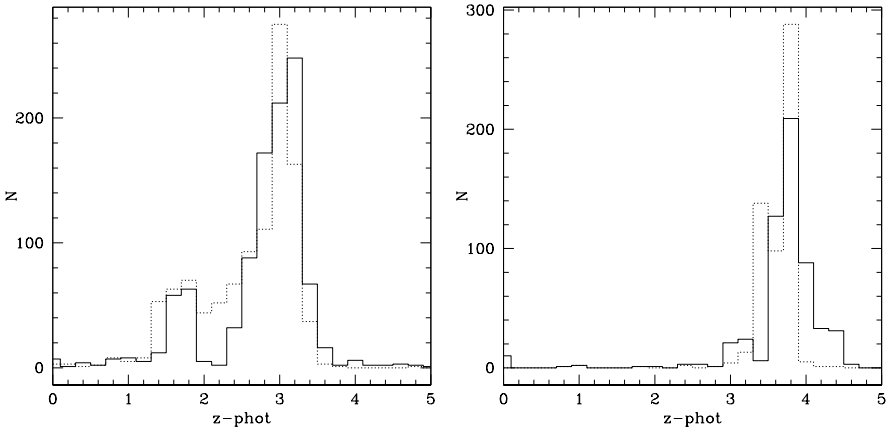}
\caption{\label{fig:z_dist}Photometric redshift distributions of
  our $U$-dropouts (\emph{left}) and our $B$-dropouts (\emph{right}). The
  solid lines correspond to the distributions of the primary solutions and the
  dashed lines correspond to the distributions of the weighted mean redshifts
  (see text).}
\end{figure*}

Photometric redshifts for all candidates are estimated from their $UBVRI$
photometry with the programme \emph{Hyperz} \citep{2000A&A...363..476B}. Again
the template SEDs by \citet{1993ApJ...405..538B} are chosen.  The programme
calculates galaxy colours in the WFI filter set at different redshifts for
every template incorporating different ages, different amounts of reddening
\citep{2000ApJ...533..682C}, and absorption by the Lyman-$\alpha$ forest
(calculated in dependence of redshift according to
\citeauthor{1995ApJ...441...18M} \citeyear{1995ApJ...441...18M}). Every object
is assigned the redshift of the best-fit SED, the primary solution phot-$z$.
Furthermore, a weighted mean redshift is computed in the 99\% confidence
interval around the primary solution. The distributions of these quantities
for all of our dropouts are shown in Fig.~\ref{fig:z_dist}. There are clear
peaks at the targeted redshifts of $z\sim3$ and $z\sim4$. Furthermore, there
is a secondary peak in the redshift distribution of the $U$-dropouts at lower
redshift ($z\sim1.7$) which is more pronounced in the distribution of the
primary redshifts and `washed-out' in the distribution of the weighted mean
redshifts.

\begin{figure*}
\includegraphics[width=17cm]{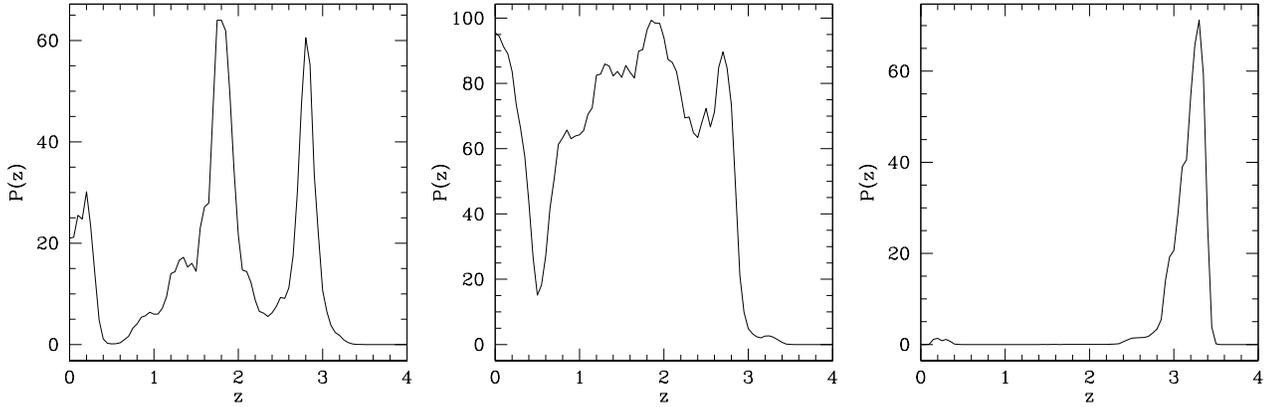}
\caption{\label{fig:log_phot}Redshift vs. probability (associated with the
  $\chi^2$ value) for three different $U$-dropouts. \emph{Left:} An object
  with assigned redshift $\mbox{phot-}z=1.81$.  This example illustrates that
  the assignment of a single number for the photometric redshift can be
  misleading. The peak at $z=2.8$ has nearly the same probability.
  \emph{Middle:} An object with assigned redshift $\mbox{phot-}z=1.85$. This
  example illustrates that sometimes the photometric redshift estimation
  totally fails but nevertheless a primary solution is put out. \emph{Right:}
  An object with assigned redshift $\mbox{phot-}z=3.29$. The ideal case of an
  object with a definite single redshift estimate.}
\end{figure*}

The programme \emph{Hyperz} can also put out the probability (associated with
the $\chi^2$ value) of an object to be located at the different redshift
values. Investigating these redshift-probability distributions of every single
$U$-dropout with assigned redshift $\mbox{phot-}z<2$ it becomes clear that for
most of them no unique solution is found. There are multiple solutions (often
another peak at $z\sim3$) or plateaus (see Fig.~\ref{fig:log_phot}), which
results in a washed-out distribution of the weighted mean redshifts. For
comparison, the unambiguous redshift probability distribution of an object
with assigned redshift $\mbox{phot-}z=3.29$ is also shown in Fig.
\ref{fig:log_phot}. Most of the objects with assigned redshift
$\mbox{phot-}z\sim3$ show similar distributions even though not as perfect as
this example.

It is possible that a significant fraction of those objects with
$\mbox{phot-}z<2$ are indeed LBGs at redshifts of 3, but are not
unambiguously identified as such by \emph{Hyperz}. This assumption is
strengthened by the fact that the redshift-probability distribution of many of
those galaxies shows a secondary peak at $z\sim3$ (see Fig.
\ref{fig:log_phot}).

In case those objects are indeed at a lower redshift of $z\sim2$, they would
fall into the so-called `redshift desert', where a determination of
photometric and spectroscopic redshifts is very difficult due to the absence
of prominent spectral features such as strong breaks in the $UBVRI$ range.

\citet{2004ApJ...604..534S} have identified a large number of galaxies in this
so-called `redshift desert' applying a technique very similar to the
Lyman-break technique with a selection box just below the LBG selection box
\citep{2004ApJ...607..226A}.  Photometric errors and differences in their and
our filter set could scatter some lower redshift objects into our selection
box. Some of these objects will certainly be included in the spectroscopic
follow-up survey described in Sect.~\ref{sec:conclusions}.

\subsection{Distribution of apparent magnitudes}
\begin{figure*}
\includegraphics[width=17cm]{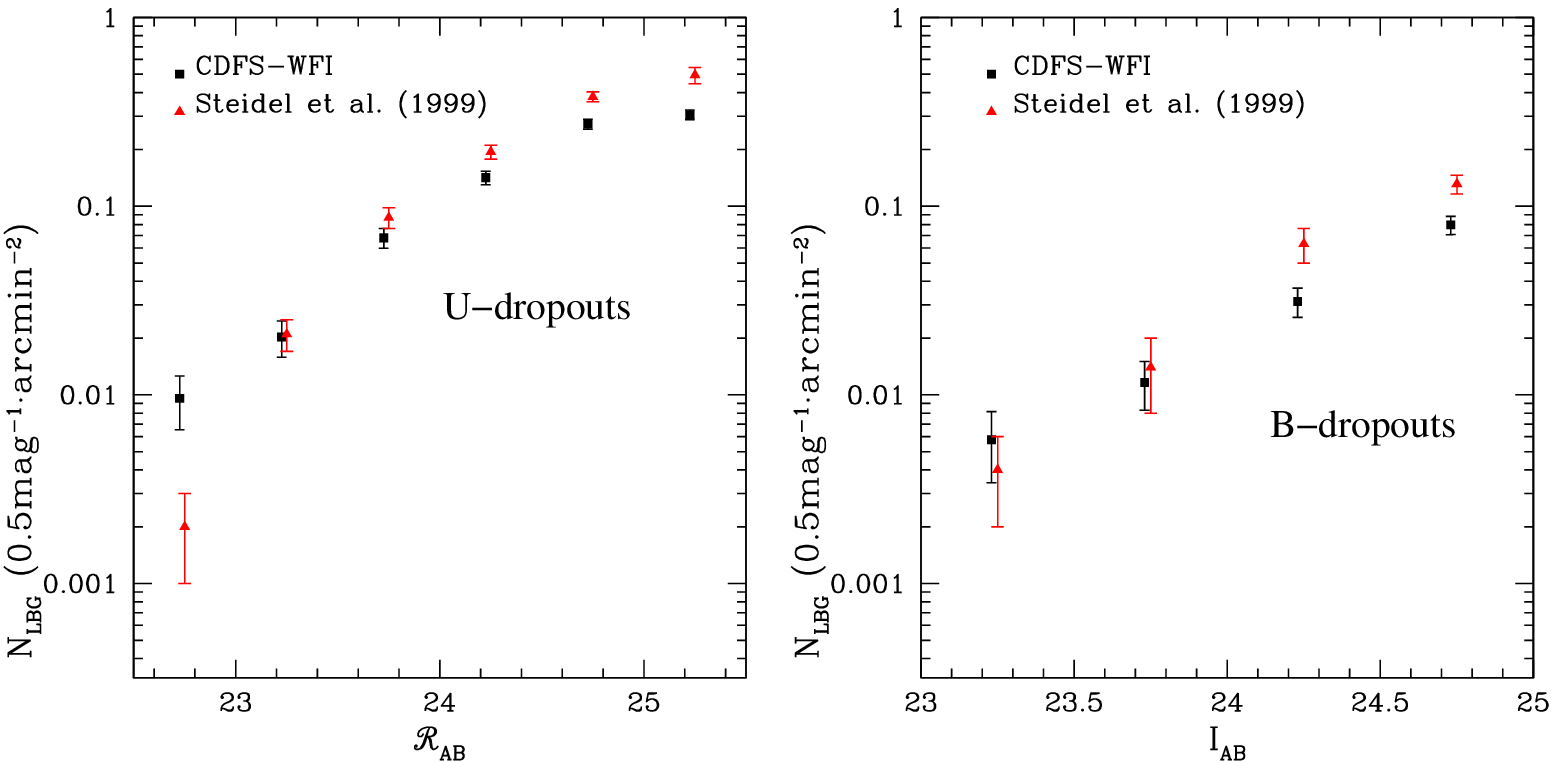}
\caption{\label{fig:number-counts}The diagrammes show number-counts for
  $U$-dropouts (\emph{left}) and $B$-dropouts (\emph{right}) of our catalogue
  (squares) and that of \citet{1999ApJ...519....1S} (triangles). The CDFS-WFI
  data points are slightly offset for clarity.}
\end{figure*}

In order to compare our number-counts to other studies, the total $R$-band
Vega magnitudes of the $U$-dropouts are converted to Steidel's
$\mathcal{R}_{AB}$-band using the transformation equation in
\citet{1993AJ....105.2017S} and an AB correction of 0.2 magnitudes.  The total
$I$-band Vega magnitudes of the $B$-dropouts are converted to the AB system
using an AB correction of 0.5 magnitudes. Both AB corrections are calculated
with \emph{Hyperz} \citep{2000A&A...363..476B}. In Fig.
\ref{fig:number-counts} our results are shown in comparison to
\citet{1999ApJ...519....1S}. In general there is good agreement between the
two studies.

The few deviations, however, can be explained. On the one hand,
\citet{1999ApJ...519....1S} correct their number-counts for contamination by
low-$z$ interlopers and stars using their large spectroscopic database which
reduces the numbers at the brighter magnitudes. On the other hand, their
images go slightly deeper ($0.5-1$~mag in the $U$-band depending on the field;
see \citeauthor{2003ApJ...592..728S}, \citeyear{2003ApJ...592..728S}) which
increases their number of LBGs at the faint end.

\subsection{Angular correlation functions}
\begin{figure*}
\includegraphics[width=17cm]{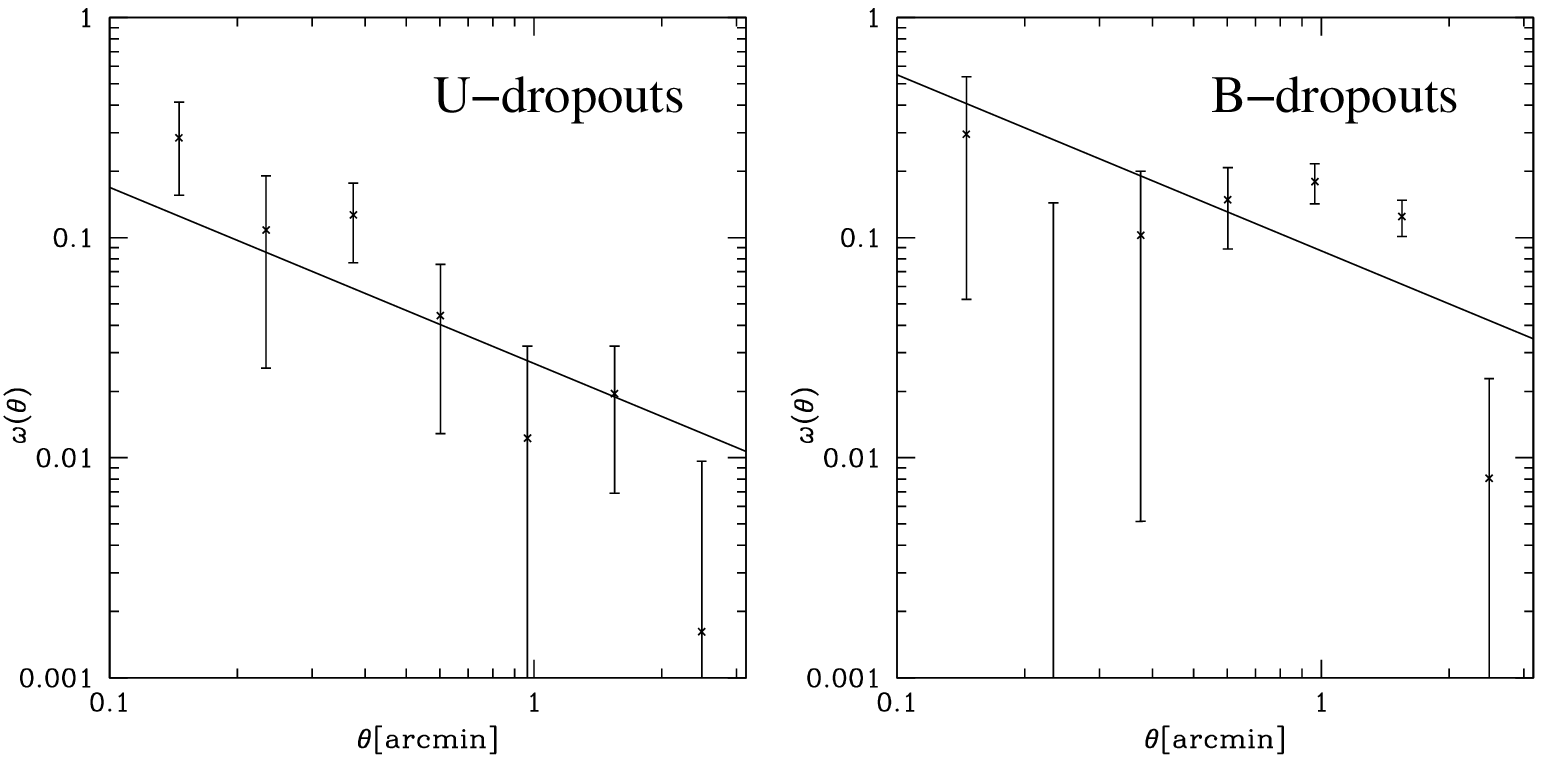}
\caption{\label{fig:correlation}Angular correlation functions for our
  $U$-dropouts (\emph{left}) and our $B$-dropouts (\emph{right}). The errors
  are Poissonian errors and the lines represent power-law $\chi^2$ fits to the
  data with a fixed slope of $\delta=0.8$. The fitted amplitude at a scale of
  $1''$ then becomes $A_\omega=0.71\pm0.13$ for the $U$-dropouts and
  $A_\omega=2.31\pm0.78$ for the $B$-dropouts.}
\end{figure*}

For calculating the angular correlation function we apply the estimator by
\citet{1993ApJ...412...64L},

\begin{eqnarray}
  \label{eqn:estimator_Landy}
  \omega(\theta)\;\delta\theta=\frac{\mathrm{DD-2DR+RR}}{\mathrm{RR}}.
\end{eqnarray}
DD, DR, and RR represent the number-counts of galaxy pairs with a separation
between $\theta$ and $\theta+\delta\theta$ in catalogues extracted from the
data (DD), from a random distribution of galaxies (RR) with the same survey
geometry (including masked out regions), and between data and random
catalogues (DR). The errors of the angular correlation function are estimated
following the Poissonian variance approach of \citet{1993ApJ...412...64L},
which is justified in the weak clustering regime,
\begin{eqnarray}
  \label{eqn:err_estimator_Landy}
  \delta\omega(\theta)=\sqrt{\frac{1+\omega(\theta)}{\mathrm{DD}}}\;.
\end{eqnarray}
A power law $\omega(\theta)=A_{\omega}\,\theta^{-\delta}$ with fixed slope
$\delta=0.8$ is fitted to the data for angular scales smaller than
$\sim2\farcm 5$. For larger scales the finite size of the fields begins to
play a role which can be accounted for by an additional constant called
`integral constraint'.  In Fig.~\ref{fig:correlation}, the angular
correlation functions for the $U$- and $B$-dropouts are shown, the latter
still suffering from small number statistics. The power law fits to our data
yield amplitudes for a scale of $1''$ of $A_{\omega}=0.71\pm0.13$ for the
$U$-dropouts and $A_{\omega}=2.31\pm0.78$ for the $B$-dropouts, respectively.

Next, we discuss the influence of inhomogeneous depth in our data on the
correlation analysis. All of our $U$-dropouts are brighter than $R=26$ and
$V=25.8$.  Given the depth of the $V$- and $R$-band image (see
Table~\ref{tab:CDFS}) small fluctuations in limiting magnitude over the field
will have no impact on our selection. In the $U$-band image, however, there
are some regions which are significantly shallower and could influence our
selection. At the left edge there is a vertical stripe and in the middle there
is a horizontal stripe where the $1\sigma$ limiting magnitude drops to
$U\sim26.4$.  Thus some objects which are faint in $V$ could be misclassified
as $U$-dropouts in these regions. For three reasons we believe that this is
not the case. First, investigating the distributions of apparent magnitudes in
the $V$- and $R$-band there is no noticeable difference between the whole
sample and the subsample in the shallower regions (288 $U$-dropouts). If
misclassification was present one would see an excess in the faint $V$-band
counts for the subsample. Second, the number density of $U$-dropouts does not
change from deep to shallow regions.  Finally, as \citet{2004ApJ...604..534S}
showed, objects that are near in colour space are mostly also near in redshift
space so that no spurious clustering signal is expected. A similar
consideration applies to the $B$-dropout sample.

For a known redshift distribution the angular correlation function
$\omega(\theta)$ can be related to the real space 3D correlation function
$\xi(r)$ using the Limber equation \citep[see][]{1980lssu.book.....P} for a
flat universe.

\begin{eqnarray}
  \omega(\theta)=\int_0^\infty \mathrm{d}\overline{w} \:
  p^2(\overline{w}) \int_{-\infty}^\infty \mathrm{d}\Delta w \:
  \xi \bigl( \sqrt{(\overline{w}\theta)^2+\Delta w^2} \bigr)\; ,
\end{eqnarray}
where $w$ is the comoving distance, and $\overline{w}$ and $\Delta w$ are the
mean and difference of the comoving distances of the two galaxies considered.
$p(\overline{w})$ is the normalised distribution of the galaxies in comoving
distance. Usually the real-space correlation function is fitted with a power
law with slope $\gamma=\delta+1$ and correlation length $r_0$:

\begin{eqnarray}
  \xi(r)=\Bigl(\frac{r}{r_0}\Bigr)^{-\gamma}\; .
\end{eqnarray}
Thus the second integral can be solved analytically, and the correlation
length $r_0$ then becomes:

\begin{eqnarray}
\label{eq:r_0}
  r_0=\left[A_{\omega;\mathrm{rad}}\cdot
  \frac{\Gamma(\gamma/2)}{\Gamma(1/2)\Gamma(\gamma/2-1/2)} \cdot
    \left(\int_0^\infty \mathrm{d}\overline{w} \:
  p^2(\overline{w})\:\overline{w}^{1-\gamma}\right)^{-1}\right]^{1/\gamma}
\end{eqnarray}
with $A_{\omega;\mathrm{rad}}$ being the amplitude of the angular correlation
function at a scale of one radian (extrapolation), and $\Gamma$ is the Euler
Gamma function.
  
In order to relate the angular correlation function to the real-space
correlation function we need to make an assumption on the redshift
distribution. For our dropout samples we choose different redshift
distributions to investigate the impact of this uncertain quantity on the
correlation lengths. First we assume flat distributions of the source
redshifts with different widths. Then we fit a Gaussian to each redshift
distribution in Fig.~\ref{fig:z_dist} neglecting the secondary peak in the
$U$-dropout redshift distribution. We find that the $U$-dropout data are well
fitted by a Gaussian with mean $z=3.03$ and a $\mathrm{FWHM}=0.54$ and the
$B$-dropout data by a Gaussian with mean $z=3.83$ and $\mathrm{FWHM}=0.34$.
The errors for the correlation lengths are estimated from the errors of the
amplitudes $A_\omega$ only, and no effects of the slope or the redshift
distributions are taken into account.  In Table
\ref{tab:clustering_measurements} the results are shown in comparison to other
studies by \citet{2004ApJ...611..685O} and \citet{2001ApJ...550..177G}. Given
the large uncertainties in our redshift distribution and the different depths
of the surveys the differences are not significant. Furthermore, within the
uncertainties, we do not see an evolution of the scale length from our
$U$-dropout sample to our $B$-dropout sample. It should be kept in mind that
the two populations do not probe the same part of the luminosity function.
With a distance modulus of $0.6\mathrm{mag}$ in a
$\Lambda\mathrm{CDM}$-cosmology between $z=3$ and $z=3.8$ and a negligible
k-correction between the $R$-band at $z=3$ and the $I$-band at
$z=3.8$\footnote{Relation between the central wavelengths (CWL):
  $\mathrm{CWL}_{R}=652\mathrm{nm}\approx\frac{(1+3)}{(1+3.8)}\cdot\mathrm{CWL}_{I}=\frac{(1+3)}{(1+3.8)}\cdot784\mathrm{nm}$}
the $U$-dropout sample is slightly deeper in terms of absolute magnitude. It
would be desirable to cut the two samples at the same $L/L^*$-value. But with
the available samples being cut at brighter magnitudes (e.g. $R\la24.6$ for
$L\ga L^{*}$ for the $U$-dropouts), the statistical errors are still too large
to reach significant conclusions. A more sophisticated clustering analysis
with dropout samples cut at the same absolute magnitude will be presented when
more fields of the DPS are available and LBG numbers have increased.

\begin{table*}
\caption{Clustering measurements. \citet{2001ApJ...550..177G} analysed a
  photometric sample  of $U$-dropouts extracted from ground-based
  data. \citet{2004ApJ...611..685O} measured the clustering on their
  $B$-dropout sample from the Subaru Deep Survey. The $I$-band limiting
  magnitude of our dropout sample is not well known since some objects are not
  detected in $I$ (see Sect.~\ref{sec:selection-candidates}).}
\label{tab:clustering_measurements}
\centering
\begin{tabular}{c c c c c}
\hline\hline
sample &  mean redshift    & redshift distribution & limiting
magnitude & $r_0$ [Mpc$\cdot h^{-1}$]    \\
\hline
$U$-dropouts (this paper)   & 3.0    & flat $2.7<z<3.3$               & $R_{\mathrm{WFI,Vega}}<26$ & $2.0\pm0.2$\\
$U$-dropouts (this paper)   & 3.0    & flat $2.6<z<3.4$               & $R_{\mathrm{WFI,Vega}}<26$ & $2.4\pm0.2$\\
$U$-dropouts (this paper)   & 3.0    & gauss $\mu=3.03$, $\sigma=0.27$& $R_{\mathrm{WFI,Vega}}<26$ & $2.6\pm0.3$\\
\citet{2001ApJ...550..177G} & 3.0    & from spectroscopic subsample & $\mathcal{R}_{AB}<25.5\hat=R_{\mathrm{WFI,Vega}}\sim25.1$& $3.2\pm0.7$\\
\hline
$B$-dropouts (this paper)   & 3.8    & flat $3.7<z<4.2$                & $I_{\mathrm{WFI,Vega}}\la26.3$ & $3.2\pm0.6$\\
$B$-dropouts (this paper)   & 3.8    & flat $3.6<z<4.3$                & $I_{\mathrm{WFI,Vega}}\la26.3$ & $3.8\pm0.7$\\
$B$-dropouts (this paper)   & 3.8    & gauss $\mu=3.83$, $\sigma=0.17$ & $I_{\mathrm{WFI,Vega}}\la26.3$ & $3.5\pm0.7$\\
\citet{2004ApJ...611..685O} & 4.0    & from simulations                & $i'_{AB}<26$ & $4.1\pm0.2$\\
\hline
\end{tabular}
%\end{minipage}
\end{table*}

\section{Conclusions and Outlook}
\label{sec:conclusions}
We find 1070 $U$- and 565 $B$-dropout candidates in deep wide-field images of
the CDFS taken with the WFI@MPG/ESO2.2m. The photometric redshift
distributions are narrowly peaked around $z=3$ and $z=4$, as expected.  Our
number-counts of dropouts in apparent magnitude bins are consistent with
previous studies. The angular correlation functions are calculated from the
data and correlation lengths are derived taking into account the photometric
redshift estimates of the samples. These results are also in good agreement
with previous studies showing no evolution from $z\sim3$ to $z\sim4$, albeit
large systematic errors remain.

The dropout samples in the CDFS will be investigated further. In Sect.
\ref{sec:selection-candidates} it was mentioned that ACS@HST images are
available for the whole WFI field. The morphology of every candidate will be
classified with the help of the high angular resolution of these data; this
will yield the largest catalogue of morphologically studied LBGs. Furthermore,
infrared data from the GOODS project \citep{2004ApJ...600L..93G} are publicly
available. The innermost part of the field (50 arcmin$^2$) is covered with
deep $JHK_s$ images from ISAAC@VLT which will help to improve the photometric
redshift accuracy considerably. A larger fraction of the area is covered with
shallower data from SOFI@NTT. For the brighter dropouts these data will also
be sufficient to improve the photometric redshift estimates.  

The aim of this study was to test techniques on the CDFS that will be applied
to a much larger dataset, the ESO Deep-Public-Survey (DPS).  This survey
covers three square degrees in total, distributed over three fields of four
adjacent WFI pointings each. Deep coverage in the $UBVRI$ bands was intended.
Unfortunately, the survey was not finished so that now there are only five
pointings (1.25 square degrees) complete in all five colours. The completion
of five further fields (that are nearly complete) was proposed by us for ESO
period 75. First investigations in the four other fields yield a number of
$U$-dropouts each comparable to the CDFS and we proposed a spectroscopic run
with VIMOS on one subfield for ESO period 76. There will be several hundreds
of LBG spectra to be analysed enabling us to quantify the contamination of our
samples, to investigate their redshift distributions, and to study the
astrophysical properties in detail. The area of 1.25 square degree that is
completely covered in all five optical bands already now yields a larger LBG
sample at $z\!\sim\!3$ than any other study to date. If the DPS is completed,
there will be $\sim\!10\,000$ $U$-dropouts in the survey on two contiguous
fields of one degree width and one of 0.5 degrees width.  From these the
clustering properties can be studied with unprecedented accuracy on the
largest scales up to now and statistics of LBG properties will be improved
significantly.

\begin{acknowledgements}
  This work was supported by the German Ministry for Education and Science
  (BMBF) through the DLR under the project 50 OR 0106, by the BMBF through
  DESY under the project 05AE2PDA/8, and by the Deutsche
  Forschungsgemeinschaft (DFG) under the project SCHN342/3--1.
\end{acknowledgements}

\bibliographystyle{aa}

\bibliography{2544_astroph}
\Online
\appendix
\section{LBG-catalogues}
\begin{longtable}{r c c r c c c c c}
\caption{\label{tab:cat_U}
  Catalogue of the $U$-dropouts in the CDFS. The magnitudes are the raw
  \emph{SExtractor} measurements not corrected for the limiting magnitudes
  (see Table~\ref{tab:CDFS}). Especially objects not detected in the $I$-band
  ($I>26.3$) may have inferior photometric redshift solutions. A photometric
  redshift value of $-9.90$ corresponds to an object for which no reasonable
  solution could be found by
  the photometric redshift code.}\\
\hline \hline
&          &          &          & \multicolumn{5}{c}{Vega mag}\\
No. & $\alpha$ (J2000.0) & $\delta$ (J2000.0) & phot-$z$ & $U$ & $B$ & $V$ & $R$ & $I$\\
\hline \endfirsthead
\caption{Catalogue of the $U$-dropouts in the CDFS}\\
\hline\hline
No. & $\alpha$ (J2000.0) & $\delta$ (J2000.0) & phot-$z$ & $U$ & $B$ & $V$ & $R$ & $I$\\
\hline
\endhead
\endfoot
  171 & 52.825630 & -28.062200 & 2.92 & 99.00 & 26.15 & 25.33 & 25.07 & 24.39\\ 
  195 & 53.039803 & -28.062126 & 3.12 & 26.70 & 25.89 & 25.62 & 25.46 & 25.54\\ 
  210 & 53.183644 & -28.062065 & 6.11 & 27.00 & 26.07 & 25.66 & 25.45 & 27.36\\ 
  237 & 53.190596 & -28.061910 & 3.31 & 99.00 & 26.36 & 25.44 & 25.16 & 25.91\\ 
  427 & 53.065301 & -28.059645 & 5.25 & 26.84 & 25.64 & 25.42 & 25.13 & 26.18\\ 
  517 & 53.105395 & -28.058580 & 2.75 & 27.42 & 25.87 & 25.21 & 25.26 & 26.02\\ 
  550 & 53.202695 & -28.057963 & 3.09 & 99.00 & 26.13 & 25.77 & 25.69 & 25.55\\ 
  587 & 53.120578 & -28.057833 & 2.88 & 26.45 & 25.81 & 25.35 & 25.16 & 24.93\\
\end{longtable}

\begin{longtable}{r c c r c c c c c}
\caption{\label{tab:cat_B}
  Catalogue of the $B$-dropouts in the CDFS. The magnitudes are the raw
  \emph{SExtractor} measurements not corrected for the limiting magnitudes
  (see Table~\ref{tab:CDFS}). Especially objects not detected in the $I$-band
  ($I>26.3$) may have inferior photometric redshift solutions. A photometric
  redshift value of $-9.90$ corresponds to an object for which no reasonable
  solution could be found by
  the photometric redshift code.}\\
\hline \hline
&          &          &          & \multicolumn{5}{c}{Vega mag}\\
No. & $\alpha$ (J2000.0) & $\delta$ (J2000.0) & phot-$z$ & $U$ & $B$ & $V$ & $R$ & $I$\\
\hline \endfirsthead
\caption{Catalogue of the $B$-dropouts in the CDFS}\\
\hline\hline
No. & $\alpha$ (J2000.0) & $\delta$ (J2000.0) & phot-$z$ & $U$ & $B$ & $V$ & $R$ & $I$\\
\hline
\endhead
\endfoot
  347 & 53.200803 & -28.060393 & 3.92 & 99.00 & 28.62 & 26.58 & 25.72 & 99.00\\ 
  797 & 53.265860 & -28.055671 & 3.79 & 27.79 & 28.33 & 27.12 & 25.99 & 25.80\\ 
 1080 & 52.853837 & -28.051275 & 4.41 & 99.00 & 99.00 & 27.12 & 25.52 & 24.67\\ 
 1217 & 52.883919 & -28.049668 & 3.86 & 99.00 & 28.66 & 26.84 & 25.71 & 25.71\\ 
 1375 & 53.384448 & -28.048522 & 3.65 & 26.75 & 28.15 & 27.00 & 25.71 & 25.48\\ 
 1513 & 53.303811 & -28.047295 & 3.88 & 99.00 & 28.16 & 26.61 & 25.58 & 99.00\\ 
 1555 & 52.840595 & -28.045446 & 4.00 & 99.00 & 28.14 & 26.39 & 25.54 & 25.71\\ 
 1774 & 52.870838 & -28.043567 & 3.04 & 26.10 & 28.12 & 25.83 & 25.15 & 24.58\\
\end{longtable}

\end{document}